\documentclass[pop,aip,preprint,amsmath,amssymb]{revtex4-1}

\usepackage{graphicx}
\usepackage{bm}
\renewcommand{\vec}[1]{\boldsymbol{#1}}
\usepackage{color}
\renewcommand{\textcolor}[2]{#2}

\newcommand{\JS}{J\"{u}ttner-Synge }
\def\aap{{\itshape Astron. Astrophys.} }
\def\apj{{\itshape Astrophys. J.} }
\def\pop{{\itshape Phys. Plasmas} }

\begin{document}

\title[PHYSICS OF PLASMAS]{Loading Relativistic Maxwell Distributions in Particle Simulations}

\author{Seiji Zenitani}
\affiliation{National Astronomical Observatory of Japan, 2-21-1 Osawa, Mitaka, Tokyo 181-8588, Japan.}
\email{seiji.zenitani@nao.ac.jp}

\begin{abstract}
Numerical algorithms to load relativistic Maxwell distributions
in particle-in-cell (PIC) and Monte-Carlo simulations are presented.
For stationary relativistic Maxwellian,
the inverse transform method and the Sobol algorithm are reviewed.
To boost particles to obtain relativistic shifted-Maxwellian,
two rejection methods are proposed in a physically transparent manner.
\textcolor{blue}{Their acceptance efficiencies are ${\approx}50\%$ for generic cases
and $100\%$ for symmetric distributions.
They can be combined with arbitrary base algorithms.}
\end{abstract}

\maketitle

\section{INTRODUCTION}

Because of an increasing demand in high-energy astrophysics,
numerical modeling of relativistic kinetic plasmas
has been growing in importance. 
To date, many simulations on relativistic kinetic processes have been performed,
such as the Rankine-Hugoniot problem across a relativistic shock \citep{gallant92},
magnetic reconnection and kinetic instabilities \citep{zeni07}
in a relativistically hot current sheet,\citep{harris,hoh66}
and the kinetic Kelvin-Helmholtz instability in a relativistic flow shear \citep{alves12}.
In these simulations,
one has to carefully set up
ultrarelativistic bulk flows and/or
relativistically hot plasmas in their rest frame.
Loading velocity distribution function,
i.e.,
initializing particle velocities by using random variables
according to a relativistic distribution function,
is essentially important.

In nonrelativistic particle simulations,
it is quite natural to begin with
a Maxwell-Boltzmann distribution (Maxwellian in short).
To load the Maxwellian,
the Box--Muller algorithm is widely used.\citep{bm58}
One can easily initialize a distribution with a bulk drift velocity,
by applying an offset to the particle velocities.

In relativistic simulations,
it is natural to begin with a relativistic Maxwellian,
also known as the J\"{u}tter-Synge distribution function.\citep{jut11,synge}
In order to load it,
perhaps the \citet{sobol76} algorithm is the most popular,
at least in Monte--Carlo simulation community.
The algorithm was formally proposed by \citet{sobol76}
in a Russian proceeding.
Its key results are outlined in \citet{pod77,pod83}.
Meanwhile,
it is not so clear how to initialize particles
according to the relativistic shifted-Maxwellian or
moving population of other distributions.
To the best of our knowledge,
the algorithms for the J\"{u}tter-Synge distribution have not been
applied to the relativistic shifted-Maxwellian.
Several alternative algorithms have been proposed.
\citet{swisdak13} applied
a rejection method for a log-concave distribution function.
\citet{melzani13} utilized
a numerical cumulative distribution function and
cylindrical transformation.

In this research note,
we describe numerical methods to load relativistic Maxwellians in particle simulations.
We first describe two base algorithms to load stationary relativistic Maxwellian,
the inverse transform method and the Sobol method.\citep{sobol76}
Next we apply the Lorentz transformation
to obtain the relativistic shifted-Maxwellian.
Simple rejection methods are proposed to deal with
the spatial part of the Lorentz transformation.
We validate the algorithms by test problems, followed by discussions.

\section{Stationary relativistic Maxwellian}

We consider relativistic Maxwell distributions
(\JS distribution\citep{jut11,synge}) in the following form,
\begin{equation}
\label{eq:JS}
f(\vec{u})d^3{u}
= \frac{N}{4\pi m^2c T K_2 (mc^2/T)} \exp \Big( -\frac{ \gamma mc^2 }{T} \Big) d^3{u},
\end{equation}
where $\vec{u}= \gamma\vec{v}$ is the spatial components of the 4-velocity,
$\vec{v}$ is the velocity,
$\gamma=[1-(\vec{v}/c)^2]^{-1/2}$ is the Lorentz factor,
$m$ is the rest mass,
$c$ is the light speed,
$T$ is the temperature, and
$K_2(x)$ is the modified Bessel function of the second kind.
The normalization constant is set such that
the number density is $N \equiv \int f(\vec{u})d^3{u}$.
Hereafter we set $m=1$ and $c=1$ for simplicity.
We use uppercases for fluid quantities and
lowercases for particle properties
throughout the paper.

To generate $\vec{u}$,
we consider the spherical transformation
$(u_x, u_y, u_z) = ( u \sin \theta \cos \varphi, u \sin \theta \sin \varphi, u \cos \theta )$.
Then Equation \ref{eq:JS} yields
\begin{equation}
\label{eq:exp_u2}
f(u) du 
=\frac{N}{T K_2 (1/T)}
\exp \Big(-\frac{\sqrt{1+u^2}}{T} \Big) u^2 du.
\end{equation}
In the special case of $N=1$,
one can read this equation
as a probability function with respect to $u$.

We generate $u$ whose distribution follows Equation \ref{eq:exp_u2}
by either the inverse transform method (Sec.~\ref{section:inverse}) or
the Sobol method (Sec.~\ref{section:sobol}).
We will describe these methods in the next subsections.

After we obtain $u$,
we generate $\vec{u}$ on a spherical surface $|\vec{u}|=u$ in the momentum space.
Using uniform random variables $X_1 (0<X_1\le 1)$ and $X_2 (0<X_2\le 1)$,
we set $\vec{u}$ in the following way,
\begin{eqnarray}
\label{eq:spherical_scattering}
\left\{ \begin{array}{lll}
u_x &=& u ~ ( 2 X_1 - 1 ) \label{eq:ux} \\
u_y &=& 2 u \sqrt{ X_1 (1-X_1) } \cos(2\pi X_2) \label{eq:uy} \\
u_z &=& 2 u \sqrt{ X_1 (1-X_1) } \sin(2\pi X_2) \label{eq:uz}
\end{array} \right.
.
\end{eqnarray}
Then we obtain a relativistic Maxwellian which follows Equation \ref{eq:JS}.

\subsection{Inverse transform method}
\label{section:inverse}

We consider the cumulative distribution function $F(u)$
with a practical upper bound $u_{\rm max}$,
\begin{align}
F(u) &= 
\Big( \int_0^{u} f(u) du \Big)
\Big( \int_0^{\infty} f(u) du \Big)^{-1} \nonumber \\
&\simeq
\Big( \int_0^{u} f(u) du \Big)
\Big( \int_0^{u_{\rm max}} f(u) du \Big)^{-1}
.
\end{align}
In the nonrelativistic limit of $T \ll 1$,
$u_{\rm max} = 5 v_{\rm th}$ is sufficient,
where $v_{\rm th}=\sqrt{2T}$ is the thermal speed. 
In the relativistically hot case of $T \gtrsim 1$,
Equation \ref{eq:exp_u2} behaves like $\propto \exp(-{u}/T)u^2$ for $u \gg 1$.
This decays slower than the nonrelativistic limit of
$\propto \exp [-(v/v_{\rm th})^2]v^2$,
and so we increase the upper bound to $u_{\rm max} = 20 T$.
We usually prepare a numerical table of $F(u)$ with $2000$ or more grid points.
Using a uniform random variable $X_3$, we compute
\begin{equation}
u=F^{-1}(X_3)
\end{equation}
by referring and interpolating the table. 

\subsection{Sobol method}
\label{section:sobol}

Let us consider the gamma distribution.
Its probability function $P(x)$ is given by
\begin{equation}
\label{eq:gamma}
P(x; a, b) = \frac{1}{b^a~{\rm Gamma}(a) } x^{a-1}e^{-x/b}
~~~~~(x\ge 0),
\end{equation}
where $a$ and $b$ are free parameters and ${\rm Gamma}(x)$ is the Gamma function. 
The gamma distribution with an integer parameter $a$ 
can be generated by
multiple random variables $X_i$'s ($0<X_i\le 1$)
in the following way,\citep{stat}
\begin{equation}
\frac{x}{b} = -\sum_{i=1}^{a} \ln X_i.
\end{equation}

\citet{sobol76} noticed that
the right hand side of Equation \ref{eq:exp_u2} is similar to
the third-order Gamma distributions,
\begin{equation}
\label{eq:Pgamma}
P(u; 3, T) = \Big( \frac{1}{2 T^{3}} \Big)\, \exp \Big(-\frac{u}{T} \Big) u^2.
\end{equation}
For a certain $T$, we initialize $u$ by using three random variables ($X_4 \dots X_6$),
\begin{equation}
\label{eq:gamma3}
\frac{u}{T} = -\ln X_4 - \ln X_5 - \ln X_6 = - \ln X_4X_5X_6
.
\end{equation}
Comparing the exponential parts in Equations \ref{eq:exp_u2} and \ref{eq:Pgamma},
we obtain a relativistic Maxwellian by the rejection method.
By using another random variable $X_7$,
we accept the particle if
\begin{equation}
\label{eq:rej}
\exp\Big( \frac{u-\sqrt{1+u^2}}{T} \Big) > X_7.
\end{equation}
Then we obtain $u$ which is distributed by Equation \ref{eq:exp_u2}.
Using Equation~\ref{eq:gamma3},
this criteria can be modified to
\begin{equation}
\sqrt{1+u^2} < \Big( u -T\ln X_7 \Big)
= - T\ln X_4X_5X_6X_7
.
\end{equation}
This leads to a simple form of
the Sobol's criterion,\citep{pod77,pod83}
\begin{equation}
\label{eq:sobol}
\eta^2-u^2 > 1
,
\end{equation}
where  $\eta=-T\ln X_4X_5X_6X_7$.
Note that $\eta$ and $u$ share
the same variables $X_4,X_5$, and $X_6$. 
\textcolor{blue}{Make sure to avoid zero in $X_4\ldots X_7$, because $\ln 0$ is undefined.}
Once Equation \ref{eq:sobol} (or Eq.~\ref{eq:rej}) is met,
we continue to the next step of the spherical scattering (Eq.~\ref{eq:spherical_scattering}). 

\begin{figure}[]
\begin{center}
\includegraphics[width={\columnwidth},clip]{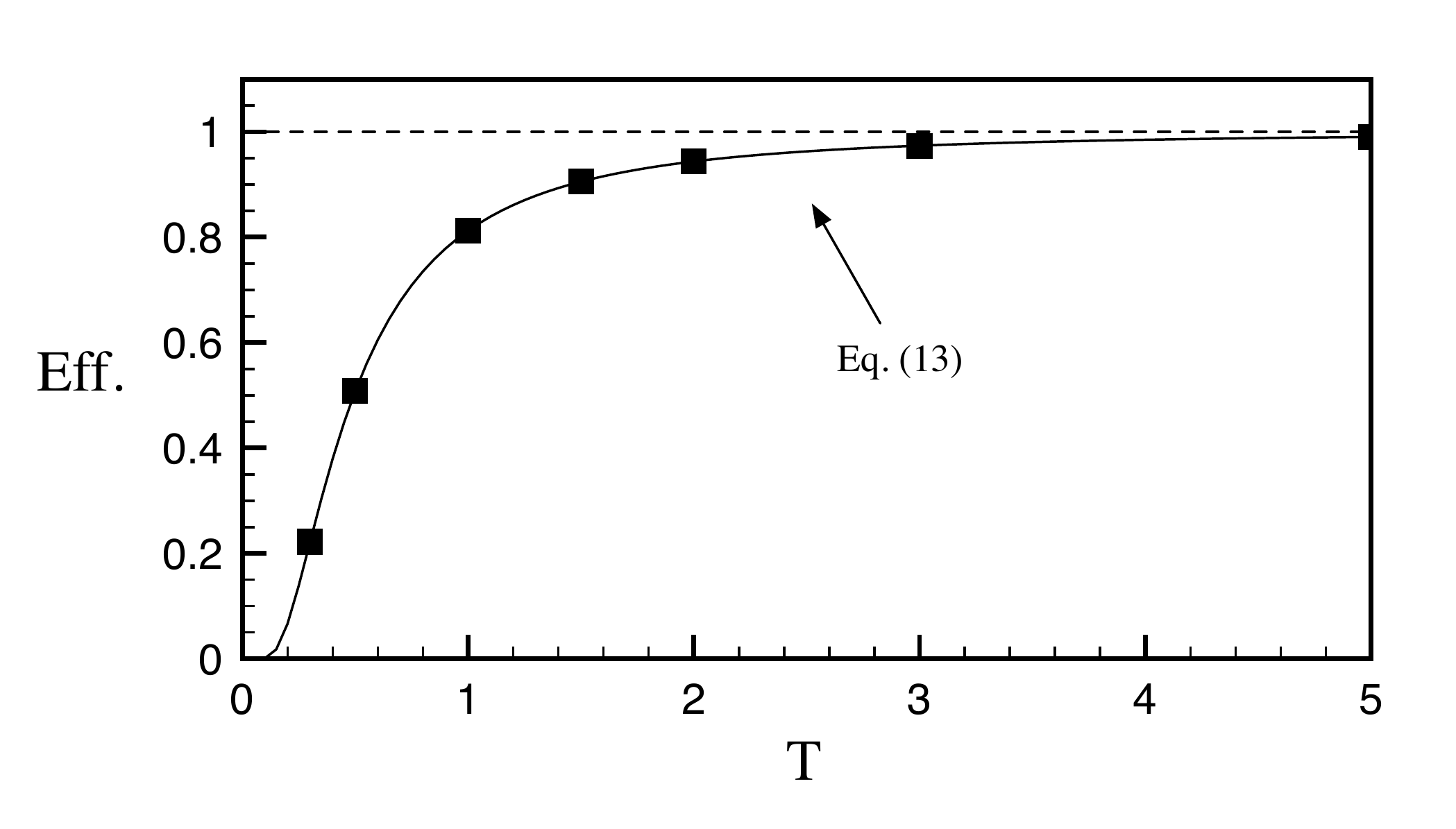}
\caption{
Acceptance efficiency of the Sobol algorithm
as a function of the temperature $T$.
The black squares show numerical results
in Section \ref{section:test}.
\label{fig:eff}}
\end{center}
\end{figure}

Comparing the normalization factors
in Equation \ref{eq:exp_u2} with $N=1$ and Equation \ref{eq:Pgamma},
we obtain the overall efficiency of the rejection method
as a function of $T$,\citep{pod77,pod83}
\begin{equation}
\label{eq:eff}
\frac{1}{2 T^2} K_2(1/T)
.
\end{equation}
Figure \ref{fig:eff} shows
the acceptance efficiency of the Sobol method,
as a function of $T$.
The efficiency quickly decreases for $T \rightarrow 0$,
while it approaches to $1$ for $T \rightarrow \infty$.

\section{Relativistic Shifted-Maxwellian}

\subsection{Lorentz transformation}

Next we discuss
general properties of the Lorentz transformation for particle distributions.
We consider the transformation between two frames, $S$ and $S'$.
We assume that particles are stationary in the reference frame $S$, and then
we switch to a moving frame $S'$ at the 4-velocity $(\Gamma,-\Gamma\beta,0,0)$.
Without losing generality, we consider the transformation in the $+x$ direction. 
In $S'$, we observe the particle distribution,
boosted by the 4-velocity $(\Gamma,+\Gamma\beta,0,0)$.
We denote the observed properties in $S'$ by the prime sign ($'$).

As the total particle number is conserved, we recognize
\begin{equation}
\label{eq:f_const}
f(\vec{x},\vec{u})\,d^3{x}\,d^3{u} = f'(\vec{x'},\vec{u'})\,d^3{x'}\,d^3{u'}
.
\end{equation}
Here, $d^3{x}=dx\,dy\,dz$ is the spatial volume element.
Using the time element $dt$ in the same frame,
we consider the 4-dimensional volume element of $dt $ and $d^3x$
that is moving at the 4-vector of $\vec{u}$.
The 4-dimensional position $(t,x,y,z)$ follows the Lorentz transformation, and so
the 4-volume element vector $(dt,dx,dy,dz)$ also follows the Lorentz transformation.
Since the Jacobian of the Lorentz transformation matrix $\Lambda$ is 1,
the 4-volume $dt\,d^3{x}$ is conserved, i.e., $dt\,d^3{x} = dt'\,d^3{x'}$. 
Since we deal with the $\vec{u}$-moving volume,
the time element $dt$ is related to the canonical time element $d\tau$
in the following way, $dt=\gamma d\tau$.
We similarly see $dt'=\gamma' d\tau$.
Therefore we obtain
\begin{equation}
\label{eq:dx}
\gamma d^3{x} = \gamma'd^3{x'}.
\end{equation}
This also indicates the length contraction for the volume. 
The transformation is slightly different for $d^3{u}$,
because $\vec{u}$ is constrained by $u^{\mu}u_{\mu} = u^2-\gamma^2 \equiv -1$. 
Without losing generality,
one can consider the Lorentz transformation
by $(\Gamma,-\Gamma\beta,0,0)$ in the $+x$ direction:
\begin{align}
\gamma' = \Gamma \gamma (1 + \beta v_x), \\
du'_x=\Gamma (du_x + \beta d\gamma), \\
du'_y=du_y, ~~~ du'_z=du_z
.
\end{align}
Under the condition of $\gamma d\gamma=u_xdu_x$,
we obtain
\begin{equation}
\label{eq:du}
\frac{d^3{u}}{\gamma} = \frac{d^3{u'}}{\gamma'}
.
\end{equation}

From Equations.~\ref{eq:dx} and \ref{eq:du},
we obtain $d^3{x}\,d^3{u}=d^3{x'}\,d^3{u'}$.
This ensures
\begin{equation}
\label{eq:f}
f(\vec{x},\vec{u})=f'(\vec{x'},\vec{u'}).
\end{equation}
We obtain a relativistic shifted-Maxwellian
by simply translating Equation~\ref{eq:f},
\begin{align}
\label{eq:shifted}
f(\vec{u}) = f'(\vec{u'})
&= \frac{N}{4\pi T K_2 (1/T)} \exp \Big( -\frac{ \gamma }{T} \Big) \\
&= \frac{N}{4\pi T K_2 (1/T)} \exp \Big( -\frac{ \Gamma(\gamma'-\beta u'_x) }{T} \Big)
\label{eq:shifted2}
\end{align}

Since we know nice algorithms (Sec.~II),
we would like to initialize the particle momentum $\vec{u}$ in the $S$ frame,
and then translate it to the $S'$ frame by the Lorentz transformation,
$\vec{u}\rightarrow\vec{u'}$.
This procedure contains the momentum-space transformation (Eq.~\ref{eq:du}).
However, it does not take care of the spatial part of the transformation,
$d^3{x}\rightarrow d^3{x'}$ (Eq.~$\ref{eq:dx}$).
Using the $S$-frame quantities,
the distribution in $S'$ appears to the observer in the following way,
\begin{equation}
f'(\vec{u'}) d^3{u'}
=
f(\vec{u}) \Big(\frac{\gamma'}{\gamma}\Big) d^3{u}
.
\end{equation}
We recognize a volume transform factor $(\gamma'/\gamma)$,
because the element volume in $S$ is not identical to the element volume in $S'$.
This issue is also addressed by \citet{melzani13}.
\begin{equation}
\label{eq:factor}
\frac{\gamma'}{\gamma} =
\Gamma ( 1 + \beta v_x )
.
\end{equation}
One can also interpret that
the number density is reciprocal to the volume size
$\propto (d^3{x})^{-1}$ (See also Eq.~\ref{eq:dx}).
Since both spacial and momentum transformation
(Equations \ref{eq:dx} and \ref{eq:du}) depends on $\vec{u}$,
the factor differs {\itshape from particle to particle}.
This may sound tricky, but
the above formula describes what the observer looks at.
We obtain very different results without the volume transformation,
as will be shown in Section IV.

In this line,
we briefly outline relativistic fluid properties.
We assume isotropic Maxwellian distribution.
From Equation \ref{eq:du},
the number flux 4-vector $N^{\mu}$ yields
\begin{equation}
\label{eq:N}
N^{\mu} = \int f(\vec{u}) u^{\mu} \frac{d^3{u}}{\gamma}
.
\end{equation}
We see $N'^{\mu} = ( N', N'\vec{V'} ) = N ( \Gamma , \Gamma \vec{\beta} )$.
Equations \ref{eq:N} ensures that
$N^{\mu}$ follows the Lorentz transformation,
i.e., $N'^{\mu} = \Lambda^{\mu}_{\alpha} N^{\alpha}$,
where $\Lambda$ is the Lorentz tensor.

Similarly, the stress-energy tensor $T^{\mu\nu}$ yields,
\begin{equation}
T^{\mu\nu} = \int f(\vec{u}) u^{\mu}u^{\nu} \frac{d^3{u}}{\gamma}
.
\end{equation}
Clearly it follows the Lorentz transformation,
i.e., $T'^{\mu\nu} = \Lambda^{\mu}_{\alpha} \Lambda^{\nu}_{\beta} T^{\alpha\beta}$.
In this case,
\begin{align}
T'^{00} &= \Gamma^2 ( \mathcal{E} + P ) - P, \\
T'^{0i} &= \Gamma^2 ( \mathcal{E} + P ) {\beta}^i,
\end{align}
where 
$\mathcal{E} \equiv \int f(\vec{u}) \gamma d^3{u} = N \{ [{K_3(1/T)}/{K_2(1/T)}]-T \}$
is the internal energy density and 
$P \equiv \int f(\vec{u})u_x ({u_x}/{\gamma}) d^3{u} = NT$
is the pressure in the rest frame.

\subsection{Volume transform methods}

Here, we describe simple methods to deal with
the volume transform factor (Eq.~\ref{eq:factor}).
It is impossible to deal with this
by adjusting the cell size in PIC simulation,
because the transformation differs from particle to particle.
One can also change the weight of particles.
However, we prefer not to do so, because
the ratio of the heaviest particle to the lightest particle could be very large.

We propose to adjust the particle number by a rejection method.
Using a random variable $X_8$ ($0<X_8\le 1$),
we accept the particle if the following condition is met,
\begin{equation}
\label{eq:factor2}
\frac{1}{2\Gamma}
\Big( \frac{\gamma'}{\gamma} \Big) = \frac{1}{2}( 1 + \beta v_x ) > X_8.
\end{equation}
The left hand side ranges from 0 to 1.
If the condition is not met, then we re-initialize the particle momentum.
The factor $(1/2\Gamma)$ can be absorbed in the normalization constant,
because we usually know the value of $2\Gamma N$ before loading particles.
The expected value ${\rm E}[x]$ of the acceptance efficiency is $50\%$,
\begin{equation}
{\rm E}\Big[ \frac{1}{2\Gamma} \Big( \frac{\gamma'}{\gamma} \Big) \Big]
= \frac{1}{2}( 1 + \beta {\rm E}[v_x] ) = 0.5.
\end{equation}
If $S$ is not the fluid rest frame,
${\rm E}[v_x] \ne 0$ and so the efficiency may vary.

\begin{figure}[]
\begin{center}
\includegraphics[width={\columnwidth},clip]{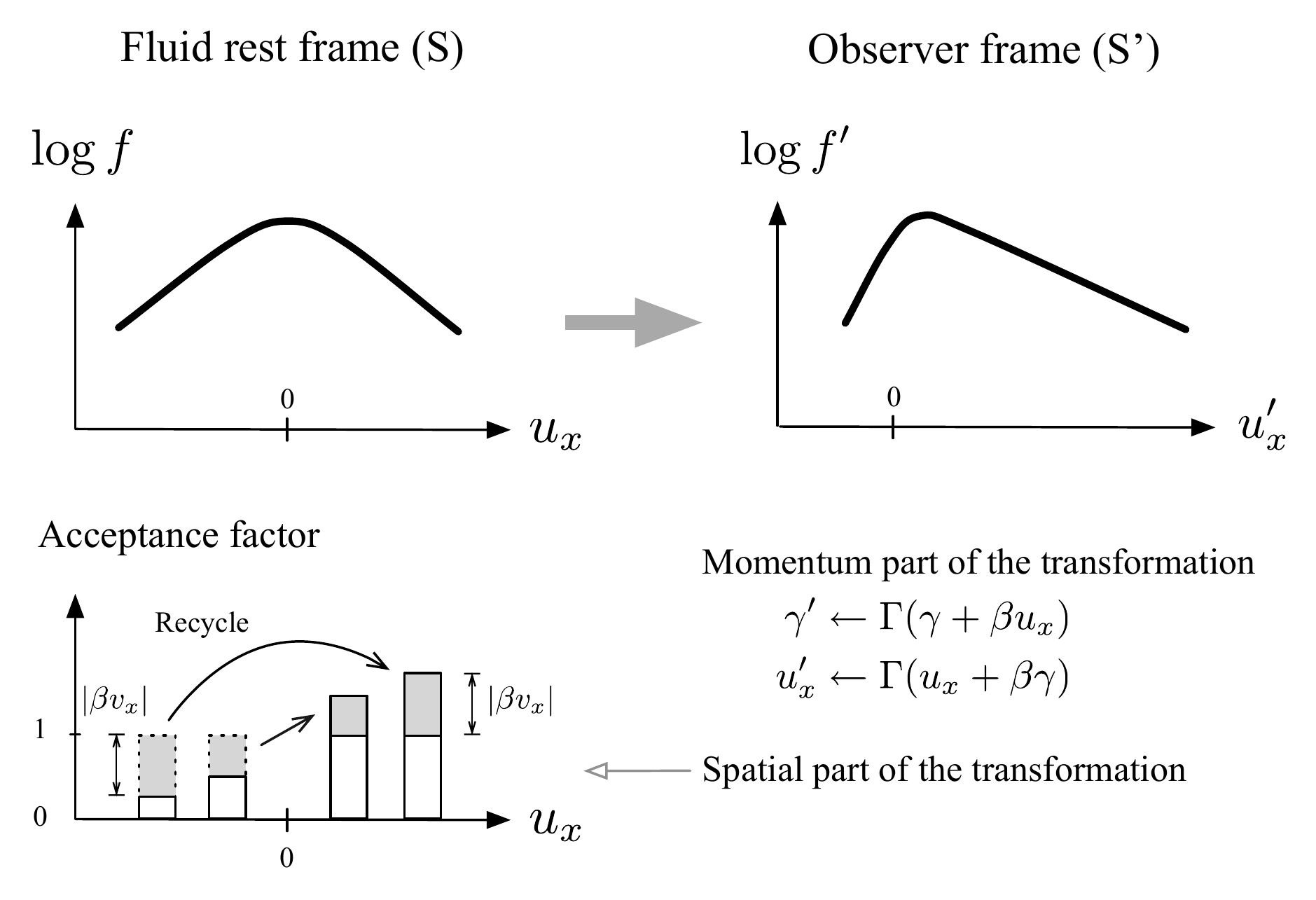}
\caption{
\textcolor{blue}{Lorentz transformation of a relativistic hot plasma distribution.
The bottom panel illustrates the flipping method,
which is responsible for the spatial part of the Lorentz transformation.}
\label{fig:shifted}}
\end{center}
\end{figure}

We can further improve
the efficiency in a special case of a symmetric distribution.
When $f(u_x)=f(-u_x)$, we multiply the acceptance factor by 2,
\begin{equation}
\frac{1}{\Gamma}
\Big( \frac{\gamma'}{\gamma} \Big) = ( 1 + \beta v_x ).
\label{eq:factor_new}
\end{equation}
The $(1/\Gamma)$ factor is absorbed in the total particle number. 
We take advantage of the fact that
the second term of the right hand side is odd function of $u_x$ (or $v_x$).
When $\beta v_x$ is negative,
the acceptance factor ranges between $0 < (1-|\beta v_x|) \le 1$.
We reject the particles at the probability of $|\beta v_x|$.
On the other hand, when $\beta v_x$ is positive,
the factor ranges between $1 \le (1+\beta v_x) < 2$.
We accept all particles.
In addition, we interpret that
we need to add another set of particles
at the probability of $|\beta v_x|$.
If $f(u_x)=f(-u_x)$,
the number of the rejected particles and
the number of the particles to be added are equal. 
We simply reverse the sign of $u_x$ of the rejected particles,
and then we add them to the positive-$\beta v_x$ side.
This logic is schematically illustrated
in the bottom of Figure \ref{fig:shifted}.


We summarize the algorithm in the following way.
If the following condition is met for a random variable $X_9$,
\begin{equation}
\label{eq:sz}
-\beta v_x > X_9,
\end{equation}
then we change $u_x \rightarrow -u_x$, before computing $u'_x$. 
Here we combine
the two conditions of $-\beta v_x < 0$ and $-\beta v_x > X_9$
to one condition (Eq.~\ref{eq:sz}). 
The acceptance efficiency is 100\%.
We call it the flipping method (Eq.~\ref{eq:sz}) to distinguish it
from the rejection method (Eq.~\ref{eq:factor2}).

\section{Test problems}
\label{section:test}

In order to validate the algorithms,
we carry out several test problems.
We initialize $10^6$ particles in all cases.
The black squares in Figure \ref{fig:eff} show
the acceptance efficiency of the Sobol method,
as a function of $T$.
They are in excellent agreement with
the expected curve (Eq.~\ref{eq:eff}).

\begin{figure}[]
\begin{center}
\includegraphics[width={\columnwidth},clip]{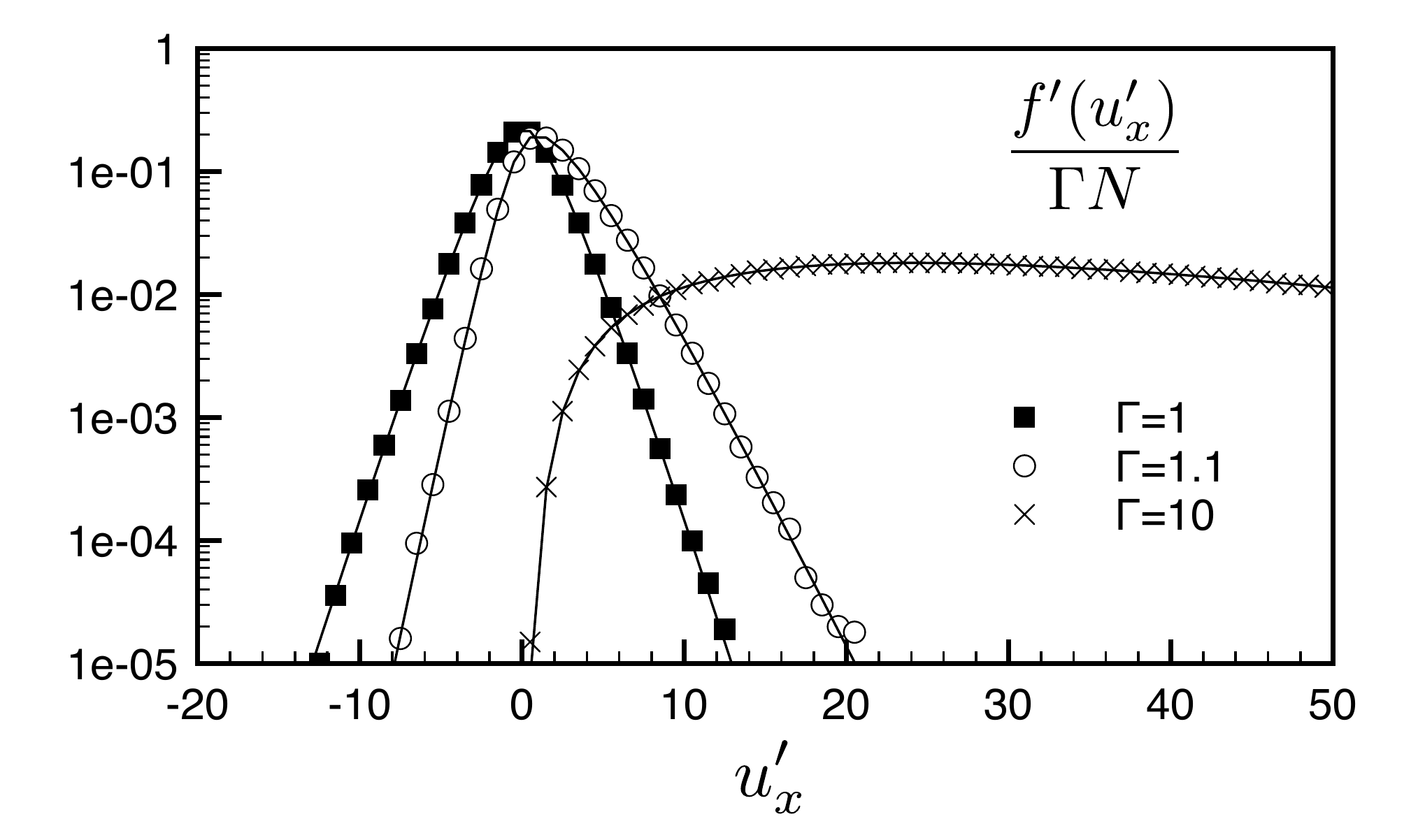}
\caption{
\textcolor{blue}{Distribution functions $f'({u}'_x)$ of
Lorentz-boosted Maxwellians as a function of $u'_x$.
Numerical results are overplotted on the analytic curves (Eq.~\ref{eq:f_ux}).
We set $T=1$ for all cases.}
\label{fig:f}}
\end{center}
\end{figure}

We then compute relativistic shifted-Maxwellian
by using the Sobol method and the flipping method (Eq.~\ref{eq:sz}).
We set $T=1$ and we boost the particles
by the bulk Lorentz factor $\Gamma = (1, 1.1, 10)$
in the $+u_x$ direction. 
\textcolor{blue}{Figure \ref{fig:f} compares
numerical results and analytic distributions
in the moving frame $S'$, integrated over $u'_y$ and $u'_z$.
All distributions are normalized by $\int f'd^3u'=\Gamma N$. 
The following analytic solution is obtained by
using a cylindrical transformation
$(u'_x, u'_y, u'_z) = ( u'_x, u'_{\perp}\cos\phi, u'_{\perp}\sin\phi )$
in Equation \ref{eq:shifted2}.
\begin{align}
f(u'_x)
&= \int^\infty_0 \int_0^{2\pi} f'(\vec{u}') u'_{\perp}  d\phi d u'_{\perp} 
\nonumber \\
&= \frac{N(\Gamma\sqrt{1+u'^2_x}+T)}{2\Gamma^2K_2(1/T)} \exp \Big( -\frac{ \Gamma(\sqrt{1+u'^2_x}-\beta u'_x) }{T} \Big)
.
\label{eq:f_ux}
\end{align}
The numerical results are in excellent agreement with the analytic solutions.
The stationary Maxwellian looks OK.
As $\Gamma$ increases,
the distribution is stretched in the $+u'_x$ direction.
From Equation \ref{eq:f_ux},
we see $f'(u'_x) \propto
u'_x \exp[ -(\Gamma(\sqrt{1+u'^2_x}-\beta u'_x)/{T}) ]
\approx u'_x\exp[ -({u'_x}/{2 \Gamma T}) ]
$ for $u'_x \rightarrow \infty$.
Therefore, the slope on the boosted side becomes extremely flat.
For $\Gamma=1.1$, 
the numerical results on the right side ($u'_x \approx 20$)
look slightly noisier than on the other side ($u'_x \approx -8$). 
This is probably a unfair comparison, because
the right slope has more gridpoints than the left slope in the low-density range. 
For $\Gamma=10$, the distribution is highly stretched in $+u'_x$.
Outside the figure, it still remains
$f'(u'_x)/\Gamma N \approx 4 \times 10^{-3}$ at $u'_x=100$. 
It will be challenging to initialize such a distribution
by a direct rejection method in the $S'$ frame,
because we have to extend the parameter domain $2\Gamma$ times longer in $+u'_x$. 
This gives us another motivation to
initialize particles in $S$ and then boost it to the $S'$ frame.}

Next, we compute several fluid quantities in the moving frame $S'$.
After initializing the particles,
we compute the flow vector $N'^{\mu}$ and the stress-energy tensor $T'^{\mu\nu}$.
Then we evaluate the average velocity $N'^{x}/N'^0=\beta$ and
the average energy flux
$T'^{0x}/N'^0=\Gamma\beta (\mathcal{E}+P)/N$.
The former is a direct indicator of the bulk motion,
and
the latter, the energy flux, plays a decisive role to the system evolution.
The results are presented in Table \ref{table}.
We change two key parameters,
the bulk Lorentz factor $\Gamma = (1.1, 10, 10^2)$ and
the relativistic temperature $T=(0.1,1,10)$.
In the $T=0.1$ case, we use the inverse transform method (Sec. II A),
because the efficiency of the Sobol method falls to $\approx 0.001$.
We also test the $T=10$ case without the volume transformation.
This incorrect case is denoted by the asterisk sign ($*$).
In Table \ref{table},
the first rows show the computed results.
The second rows show the relative error to analytic solutions.
As can be seen, the results appear to be accurate,
except for the rightmost columns.

Without the volume transformation, we see that
the average bulk speed is inaccurate in Table \ref{table}.
This is crucial to initialize a relativistic current sheet \citep{harris,hoh66},
in which relativistically hot populations carry the electric current.
The energy flux is significantly distorted, too.
The average energy flux without the volume transformation is
\begin{equation}
\frac{
\int f(\vec{u}) u'^0 ({u'^x}/{u'^0}) {d^3{u}}
}
{\int f(\vec{u}) {d^3{u}}}
=
\frac{1}{N} \int f(\vec{u}) \Gamma ( \beta u^0 + u^x ) {d^3{u}}
= \Gamma \beta \frac{ \mathcal{E} }{ N }
.
\end{equation}
Since $(\mathcal{E}+P)/\mathcal{E} \rightarrow 4/3$ for $T \gg 1$,
we lose $25\%$ of the energy flux,
regardless of the bulk speed $\beta$.
We can similarly evaluate the average energy density without the volume transformation.
It deviates from the right value by a factor of
$[ 1 + \frac{\Gamma^2-1}{\Gamma^2} \big( \frac{P}{\mathcal{E}} \big) ]^{-1}$,
and therefore the error approaches $25\%$ for $\Gamma \gg 1$ and $T \gg 1$.


\begin{table}
\begin{center}
\caption{
Computed fluid quantities and relative errors.
\label{table}}
\begin{tabular}{lrrrr}
\hline
$\frac{N'^x}{\Gamma N}$ & 
T=0.1 &
T=1.0 &
T=10 &
T=10*
\\
\hline
$\Gamma$=1.1 &
0.416532	& 0.416708	& 0.416566	& 0.288896 \\
~ &
$1.6 \times 10^{-4}$	& $2.6 \times 10^{-4}$	& $7.7 \times 10^{-5}$	& 0.307 \\
$\Gamma$=10 & 
0.994989	& 0.994996	& 0.994958	& 0.975918 \\
~ &
$1.6 \times 10^{-6}$	& $8.9 \times 10^{-6}$	& $2.9 \times 10^{-5}$	& 0.0192 \\
$\Gamma$=100 & 
0.999950	& 0.999950	& 0.999950	& 0.999658 \\
~ &
$9.3 \times 10^{-10}$	& ~$4.0 \times 10^{-8}$	& ~$1.3 \times 10^{-7}$	& ~$2.9 \times 10^{-4}$ \\
\hline
$\frac{T'^{0x}}{\Gamma N}$ & 
T=0.1 &
T=1.0 &
T=10 &
T=10*
\\
\hline
$\Gamma$=1.1 &
0.580438	& 2.00167	& 18.3798	& 13.7357 \\
~ &
$2.9 \times 10^{-4}$	& $5.5 \times 10^{-4}$	& $1.4 \times 10^{-3}$	& 0.252 \\
$\Gamma$=10 & 
12.6063	& 43.4805	& 398.147	& 298.838 \\
~ &
$2.6 \times 10^{-6}$	& $1.1 \times 10^{-4}$	& $8.5 \times 10^{-4}$	& 0.250 \\
$\Gamma$=100 & 
126.674	& 437.062	& 4001.75	& 3003.43 \\
~ &
$1.5 \times 10^{-4}$	& $9.2 \times 10^{-5}$	& $7.4 \times 10^{-4}$	& 0.250 \\
\hline
\end{tabular}
\end{center}
\end{table}

\section{Discussion and Summary}

We first reviewed
two algorithms to initialize
the stationary relativistic Maxwellian.
In addition to the simple inverse transform method,
we have formally reviewed the Sobol algorithm.
In our experience,
the inverse transform method is faster than the Sobol method,
because it only requires 3 random variables.
We don't see any problems, as long as we prepare $10^3$-$10^4$ grids in the table.
The algorithm can deal with any spherically-symmetric distributions.
On the other hand, the Sobol method has a strong mathematical background.
It is very simple, and so we can easily avoid a bug.
The method is certainly slower than the inverse transform method,
because it uses 6 random variables.
However, this will not be a big deal,
because we use these algorithms for initialization. 
The only problem is that
the Sobol method becomes extremely inefficient
for the nonrelativistic limit of $T \ll 1$.
In such a case, we simply switch from the Sobol method
to the inverse transform method or the Box-Muller method.
Another promising option is the log-concave rejection method,
described in Section II and Appendix in \citet{swisdak13}.
The algorithm uses 4 random variables,
its acceptance efficiency is ${\approx}90\%$, and
it is nearly insensitive to $T$.

After initializing the stationary Maxwellian,
we apply the volume transformation before boosting the particle momentum.
We have proposed the two algorithms,
the rejection method (Eq.~\ref{eq:factor2}) and
the flipping method (Eq.~\ref{eq:sz}).
They require one more random variable.
The flipping method is our first choice.
Since it accepts all particles,
the overall efficiency is the same as
the base algorithm for the stationary one.
As a representative case,
the Sobol method with the flipping method are summarized
in the pseudocode in Table \ref{table:code}.
We emphasize that our volume transform methods are generic.
The flipping method can be combined with
power-law, waterbag, or any other distributions,
as long as it is symmetric in $u_x$ in the $S$ frame.
Even when the distribution is non-symmetric,
we can divert to the rejection method (Eq.~\ref{eq:factor2}).
The acceptance efficiency is typically $50\%$,
but it works in any cases. 
\citet{swisdak13} used the log-concave rejection method twice
for the shifted Maxwellian.
According to his article, the overall efficiency is $\approx 80\%$ insensitive to $T$.
This is a very good result.
However, his algorithm is specialized for
Maxwellians or possibly other exponential-type distributions.
In contrast, our simple methods can deal with any kind of distributions.

Using the test problems, we have demonstrated that
the combinations of the base methods and the flipping method excellently work.
Without the volume transformation,
we recognize significant errors up to $25\%$ in the average energy flux.
This is because the volume transform factor (Eq.~\ref{eq:factor}) is
no longer constant for $\Gamma > 1$ and $T \gg 1$.

In summary, we have described numerical algorithms
to load relativistic Maxwellians in particle simulations.
The inverse transform method and the Sobol method are useful
to load the stationary Maxwellian.
For shifted Maxwellian,
the rejection method (Eq.~\ref{eq:factor2}) and
the flipping method (Eq.~\ref{eq:sz}) take care of
the spatial part of the Lorentz transformation.
These methods are simple and physically-transparent.
They can be combined with arbitrary base algorithms.
We hope that these algorithms are useful
in relativistic kinetic simulations in high-energy astrophysics.

\begin{table}
\begin{center}
\caption{Sobol algorithm with the flipping method.
\label{table:code}}
\begin{tabular}{l}
\\
\hline
{\bf repeat}\\
$~~~~$generate $X_1, X_2, X_3, X_4$, uniform on (0, 1]\\
$~~~~u \leftarrow -T \ln X_1X_2X_3$\\
$~~~~\eta \leftarrow -T \ln X_1X_2X_3X_4$\\
{\bf until} $\eta^2 - u^2 > 1$.\\
generate $X_5, X_6, X_7$, uniform on [0, 1]\\
$u_x \leftarrow u ~ ( 2 X_5 - 1 )$ \\
$u_y \leftarrow 2 u \sqrt{ X_5 (1-X_5) } \cos(2\pi X_6)$ \\
$u_z \leftarrow 2 u \sqrt{ X_5 (1-X_5) } \sin(2\pi X_6)$ \\
{\bf if} ($-\beta v_x > X_7$), $u_x \leftarrow -u_x$\\
$u_x \leftarrow \Gamma (u_x + \beta \sqrt{1+u^2})$ \\
{\bf return} $u_x, u_y, u_z$\\
\hline
\end{tabular}
\end{center}
\end{table}

\begin{acknowledgements}
The author acknowledges
M. Oka and A. Taktakishvili for their assistance to find out Sobol's original article and
T. N. Kato for his insightful comments on the manuscript.
This work was supported by Grant-in-Aid for Young Scientists (B) (Grant No. 25871054).
\end{acknowledgements}


\begin{thebibliography}{}
\bibitem[Alves et al.(2012)]{alves12}
Alves, E. P., Grismayer, T., Martins, S. F., Fi\'{u}za, F., Fonseca, R. A., and Silva, L. O.,
``Large-scale Magnetic Field Generation via the Kinetic Kelvin-Helmholtz Instability in Unmagnetized Scenarios,'' \apj {\bf 746}, L14 (2012).
\bibitem[Box \& Muller(1958)]{bm58}
Box, G. E. P. and Muller, M. E., ``A note on the generation of random normal deviates,'' {\it Annals of Mathematical Statistics } {\bf 29}, 610 (1958).
\bibitem[Gallant et al.(1992)]{gallant92} Gallant, Y. A., Hoshino, M., Langdon, A. B., Arons, J., and Claire, C. E., ``Relativistic, perpendicular shocks in electron-positron plasmas,'' \apj {\bf 391}, 73 (1992).
\bibitem[Harris(1962)]{harris}
Harris, E. G., ``On a plasma sheath separating regions of oppositely directed magnetic field,'' {\it Nuovo Cimento } {\bf 23}, 115 (1962).
\bibitem[Hoh(1966)]{hoh66} Hoh, F. C., ``Stability of Sheet Pinch,'' {\itshape Phys. Fluids } {\bf 9}, 277 (1966).
\bibitem[J{\"u}ttner(1911)]{jut11} J{\"u}ttner, F., ``Das Maxwellsche Gesetz der Geschwindigkeitsverteilung in der Relativtheorie,'' {\it Ann. Phys. }{\bf 339}, 856 (1911).
\bibitem[Kennedy \& Gentle(1980)]{stat} Kennedy, W. J., Jr. and Gentle, J. E., {\it ``Statistical Computing,''} Marcel Dekker Inc. (1980).
\bibitem[Landau \& Lifshitz(1975)]{book2} Landau, L. D. and Lifshitz, E. M., {\itshape ``The classical theory of fields,''} Oxford, 4th ed, \S 6, \S 10 (1975).
\bibitem[Melzani et al.(2013)]{melzani13}
Melzani, M., Winisdoerffer, C., Walder, R., Folini, D., Favre, J. M., Krastanov, S., and Messmer P., ``Apar-T: code, validation, and physical interpretation of particle-in-cell results,'' \aap {\bf 558}, A133 (2013).
\bibitem[Pozdnyakov et al.(1977)]{pod77}
Pozdnyakov, L. A., Sobol, I. M., and Sunyaev, R. A., ``Effect of multiple Compton scatterings on an X-ray emission spectrum - Calculations by the Monte Carlo method,'' {\it Soviet Astronomy } {\bf 21} 708 (Translation) (1977).
\bibitem[Pozdnyakov et al.(1983)]{pod83}
Pozdnyakov, L. A., Sobol, I. M., and Sunyaev, R. A., ``Comptonization and the shaping of X-ray source spectra - Monte Carlo calculations,'' {\it Astrophys. Space Phys. Rev. } {\bf 2}, 189 (Translation) (1983).
\bibitem[Sobol(1976)]{sobol76} Sobol, I. M., ``On Modeling Certain Distributions Similar to Gamma Distribution,'' in Monte Carlo Methods in Computational Mathematics and Mathematical Physics (Novosibirsk, 1976), pp. 24--29 [in Russian] (1976).

\bibitem[Swisdak(2013)]{swisdak13} Swisdak, M., ``The generation of random variates from a relativistic Maxwellian distribution,'' \pop {\bf 20}, 062110 (2013).
\bibitem[Synge(1957)]{synge} Synge, J. L., {\it The Relativistic Gas}, New York: Interscience (1957).
\bibitem[Zenitani \& Hoshino(2007)]{zeni07} Zenitani, S. and Hoshino, M., ``Particle Acceleration and Magnetic Dissipation in Relativistic Current Sheet of Pair Plasmas,'' \apj {\bf 670}, 702 (2007).

\end{thebibliography}
\end{document}